\documentstyle[preprint,prc,aps]{revtex}
\tightenlines
\begin{document}
\draft
\preprint{IFA/SP-98/1}

\title{Amplitudes and Resonances from an Energy-Dependent Analysis of $\overline{p}%
+p\rightarrow \pi +\pi $ }
\author{B. R. Martin}
\address{Department of Physics and Astronomy, University College London, \\
London WC1E 6BT, England}
\author{G. C. Oades}
\address{Institute of Physics and Astronomy, Aarhus University,\\
DK-8000 Aarhus C, Denmark}
\date{submitted to Phys. Rev. C: Brief Reports}
\maketitle

\begin{abstract}
The amplitudes at a series of discrete energies obtained from a previous
analysis of $\overline{p}p\rightarrow \pi \pi $ have been used as input to a
global energy-dependent analysis of data in the momentum range 360 - 1550
MeV/c. The results confirm the previous analysis and yield refined values
for meson resonance parameters in this energy region.
\end{abstract}


\section{Data Analysis}

In a previous paper \cite{ref1} we presented the results of an analysis of
data on the reaction $\overline{p}p\rightarrow \pi \pi $ carried out at a
series of discrete energies in the center-of-mass range 1.91 MeV to 2.27
MeV. The data consisted of differential cross-sections (dcs) for both the $%
\pi ^{-}\pi ^{+}$ and $\pi ^{0}\pi ^{0}$ channels and angular asymmetry
distributions (polarizations) for the $\pi ^{-}\pi ^{+}$ channel alone.
These were supplemented by invariant amplitudes at each energy obtained in
an earlier analysis using hyperbolic dispersion relations \cite{ref2}. The
latter allowed analyticity and crossing symmetry to be imposed on the
solutions and thus ensured that they were consistent with the wealth of data
in the $\pi N\rightarrow \pi N$ channel. The amplitude constraints, used for
the first time in Ref. \cite{ref1}, enabled other published solutions to be
ruled out and produced a set of resonance parameters more reliable than
those of earlier analyses \cite{ref3}. Single-energy analyses, however, do
not include the correlations between amplitudes at different energies, nor
in our method of extracting resonance parameters did we include correlations
between different partial waves. In this paper we have therefore used the
output amplitudes from Ref. \cite{ref1} as starting values for an
energy-dependent analysis in the same energy region.

Initially we used the same data set as was used in Ref. \cite{ref1}, i.e.
dcs and polarization values at 20 momenta for the $\pi ^{-}\pi ^{+}$ channel
from Ref. \cite{ref4} (1973 data points) and dcs values at 14 momenta for
the $\pi ^{0}\pi ^{0}$ channel from Ref. \cite{ref5} (551 data points),
giving a total of 2524 data points. In the present case, since we are making
an energy dependent analysis, the latter did not have to be interpolated to
the same momenta as the measured $\pi ^{-}\pi ^{+}$ cross-sections. As in
Ref. \cite{ref1}, these experimental data were supplemented by values of the
invariant amplitudes for the annihilation channel obtained via hyperbolic
dispersion relations in Ref. \cite{ref2} (3304 data points). We have also
explored the compatibility of earlier data with this data set by including
dcs values for the $\pi ^{-}\pi ^{+}$ channel from Ref. \cite{ref6} (998
data points).

\section{Amplitudes and Resonances}

The parametrization used was the same as that used in Ref. \cite{ref1}, ie
we work in the JL basis and each partial-wave helicity amplitude for a given
J and L was written as 
\begin{equation}
h_{J\pm }(W)=\frac{\alpha _{J\pm }}{M_{R}-W-i\Gamma /2}+k^{L+1/2}%
\sum_{n=1}^{n_{JL}}\beta _{J\pm }^{(n)}x^{n-1},  \label{eq1}
\end{equation}
where $h_{J\pm }\equiv h_{J,L=J\pm 1}$. Here $W$ is the center-of-mass
energy and 
\begin{equation}
x\equiv \frac{2W-W_{\min }-W_{\max }}{W_{\max }-W_{\min }}.  \label{eq2}
\end{equation}
In the second (background) term the coefficients $\beta _{J\pm }^{(n)}$ are
complex parameters and to ensure the correct behaviour at the $\overline{N}N$
threshold we set $k=p/p_{B}$ where $p_{B}$ is the momentum corresponding to $%
W=2.1$ GeV. In the resonance term, the parameters are the mass $M_{R}$, the
width $\Gamma $ and the complex residues $\alpha _{J\pm }$. To ensure the
correct threshold behaviour at the $\overline{N}N$ threshold, we set 
\begin{equation}
\alpha _{J\pm }=\gamma _{J\pm }(\frac{p}{p_{R}})^{L+1/2},\;\;\;p\leq p_{R}
\label{eq3}
\end{equation}
\begin{equation}
=\gamma _{J\pm },\;\;\;\;\;\;\;\;\;\;\;\;\;\;\;p>p_{R}
\end{equation}
where $p_{R}$ is the value of $p$ at $W=M_{R}$ and $\gamma _{\pm }$ is a
complex constant. From the values of $\gamma _{\pm }$ and $\Gamma $ one can
calculate the product of branching ratios $B_{J}\equiv B(R\rightarrow \pi
\pi )B(R\rightarrow \overline{N}N)$. The amplitudes $h_{J\pm }$ are
dimensionless and are normalized so that the integrated cross-section for a
given isospin is given by 
\begin{equation}
\sigma =\frac{\pi }{p^{2}}\sum_{J}\left( 2J+1\right) \left\{ \left|
h_{J+}\right| ^{2}+\left| h_{J-}\right| ^{2}\right\} .  \label{eq4}
\end{equation}

The quality of fits to data over a range of energies is always considerably
worse than that obtained at a single energy. This is partly due to
normalization differences between different experiments and even between
different energies for the same experiment. Also, isolated discrepant points
may make an anomalously large contribution to whatever measure is used to
judge the quality of the fit. To reduce the latter effects, we have used
robust estimation, minimising the quantity 
\begin{equation}
\sum_{i=1}^{N_{pts}}\ln (1+0.5z_{i}^{2})  \label{eq5}
\end{equation}
where $N_{pts}$ is the total number of data points and where, for a given
data point 
\begin{equation}
z_{i}=\frac{y_{i}-y_{param,i}}{\sigma _{i}}  \label{eq6}
\end{equation}
$y_{i}$ being the input data point, $y_{param,i}$ being the prediction for
the same data point from the parametrization and $\sigma _{i}$ being the
error on the data point. This reduces the influence of isolated discrepant
points compared to the usual $\chi ^{2}$ minimization , although we will
also quote the resulting $\chi ^{2}$ values. In each fit we used as starting
values the amplitudes found in Ref. \cite{ref1}, with the resonance
parameters loosely constrained to lie close to their initial values,
typically within 50 MeV, although this was not an absolute constraint. The
point here is that we are not attempting to make a systematic search of the
entire parameter space, but rather to test the compatibility of the energy
dependence of our previous solution with the whole data set. In practice,
the resonance parameters showed no significant tendency to move away from
their initial values. Starting with the data from Refs \cite{ref4} and \cite
{ref5}, we found that a solution could be found with a $\chi ^{2}$ per data
point averaged over experiments as follows: 2.05 ($\pi ^{-}\pi ^{+}$
channel), 0.61 ($\pi ^{0}\pi ^{0}$ channel) and 0.11 (invariant amplitudes).
These values are, as expected, higher than obtained in single-energy
analyses. Allowing small renormalizations (typically less than 10\%) on the
experimental dcs data reduced these values by 21\% ($\pi ^{-}\pi ^{+}$) and
15\% ($\pi ^{0}\pi ^{0}$). In addition 69 of the 1973 $\pi ^{-}\pi ^{+}$
data points contribute more than 10 to $\chi ^{2}$; removing these reduces
the $\pi ^{-}\pi ^{+}$ average $\chi ^{2}$ per data point from 2.05 to 1.45.
However, whether or not these various adjustments are made, the resulting
solution remains essentially unchanged. We also tested the compatibility of
earlier data \cite{ref6} with the accurate dcs data from LEAR \cite{ref4} by
including the former in the fit. There was little change in either the
quality of the fit, the amplitudes or the resonance parameters and so we
conclude that the newer LEAR data \cite{ref4} are compatible with the older
dcs.

The resonance parameters of the solutions found are shown in Table \ref
{table1}. The range of values spans those found in different solutions and
using the two data sets. The parameters are rather similar to those found in
Ref. \cite{ref1}, as is the pattern of their couplings to the different
helicity states. This is discussed in detail in Ref. \cite{ref1} and will
not be repeated here. For $J=0$, the width has increased slightly and the
value of $B_{0}$ is somewhat smaller, but there is still strong evidence for
a state with an abnormally large coupling to the $\overline{N}N$ channel.
Two places where the present solution distinguishes different possibilities
found in Ref. \cite{ref1} are the $J=3$ and $J=4$ waves where in both cases
smaller couplings are preferred. In addition, for $J=4$ a smaller width is
found, closer to the width of the established $f_{4}(2050)$ although
somewhat smaller than the accepted value. The corresponding amplitudes are
shown in Fig. \ref{fig1} for solutions with and without the older $\pi
^{-}\pi ^{+}$ experiments \cite{ref6}.

\newpage 
\begin{figure}[tbp]
\caption{Partial wave helicity amplitudes in the $JL$ basis
obtained from an energy-dependent fit to the data from \protect\cite
{ref4,ref5} (solid lines) and including the older data \protect\cite{ref6}
(dashed lines). In each case the start of the argand diagram is indicated by
S.\newline
(a) $J=0,1,2.$\newline
(b) $J=3,4,5.$}
\label{fig1}
\end{figure}

\newpage 
\begin{table}[tbp]
\caption{Resonance masses and widths in units of GeV, obtained from fitting
dcs and polarization data \protect\cite{ref4,ref5,ref6}, together with the
values of the product of branching ratios .}
\label{table1}%
\begin{tabular}{dddd}
$J$ & Mass & Width & $B_J$ \\ 
\hline
0 & 1.95 & 0.17 - 0.18 & 0.13 - 0.15 \\ 
1 & 1.96 & 0.15 - 0.17 & 0.059 - 0.064 \\ 
2 & 1.93 & 0.14 - 0.15 & 0.011 \\ 
3 & 2.02 & 0.23 & 0.002 - 0.006\tablenotemark[1] \\ 
4 & 2.00 & 0.16 - 0.18 & 0.0022 - 0.0024 \\ 
5 & 2.19 & 0.22 & 0.0011 - 0.0018 
\end{tabular}
\tablenotetext[1]{There is a misprint in the corresponding entry in Table II
of \cite{ref1}, where this number is given as 0.028 instead of 0.0028.}
\end{table}

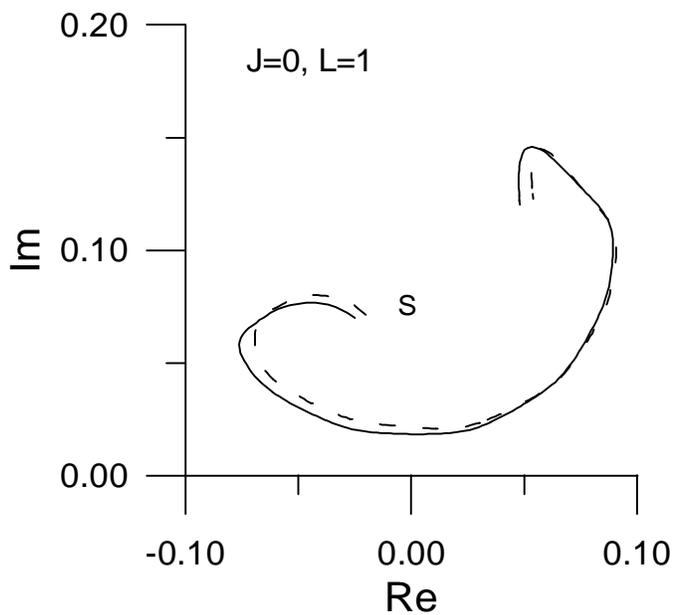
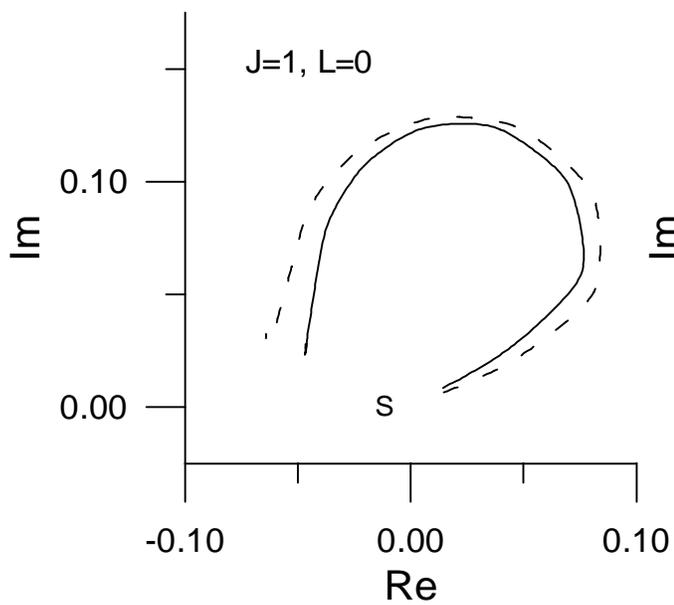
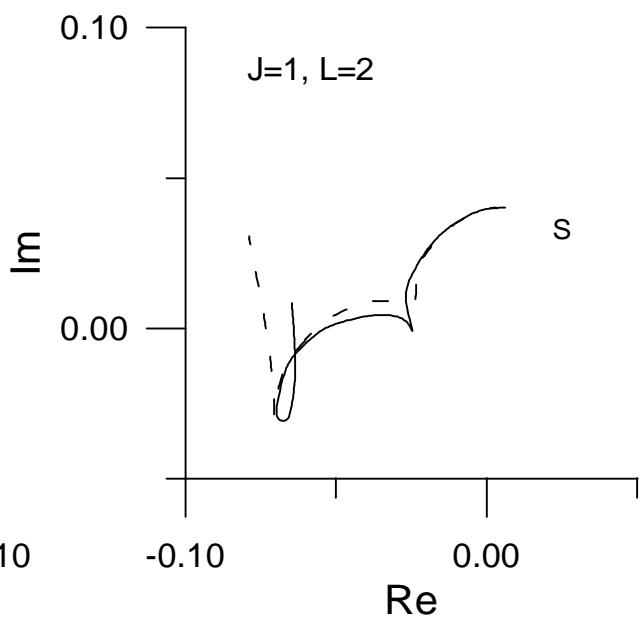
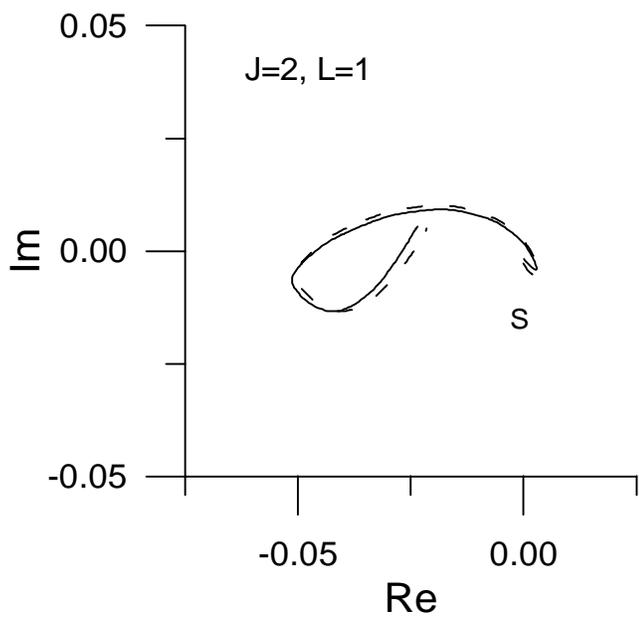
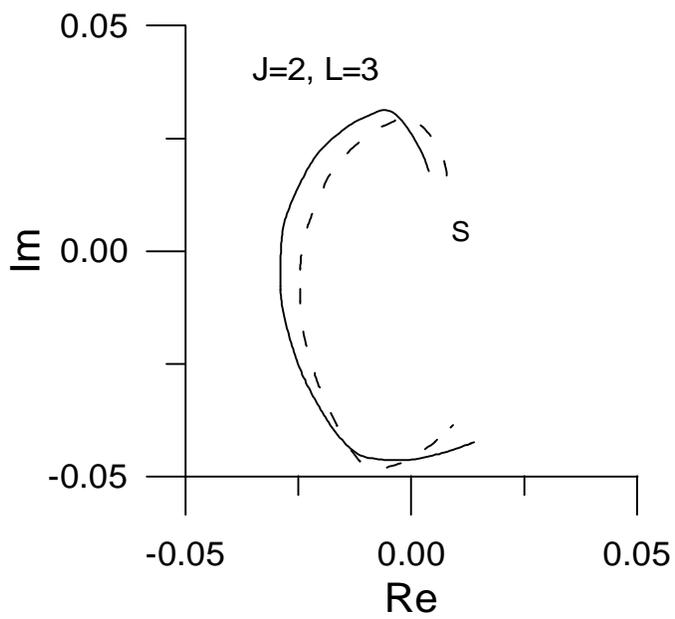
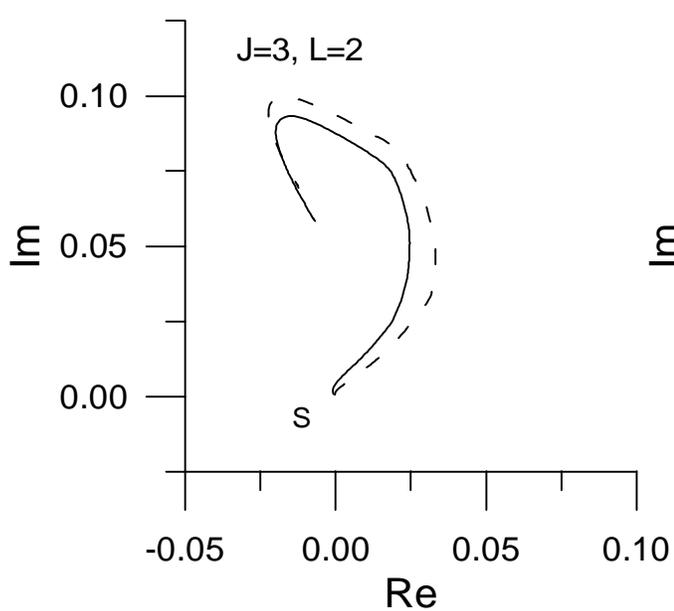
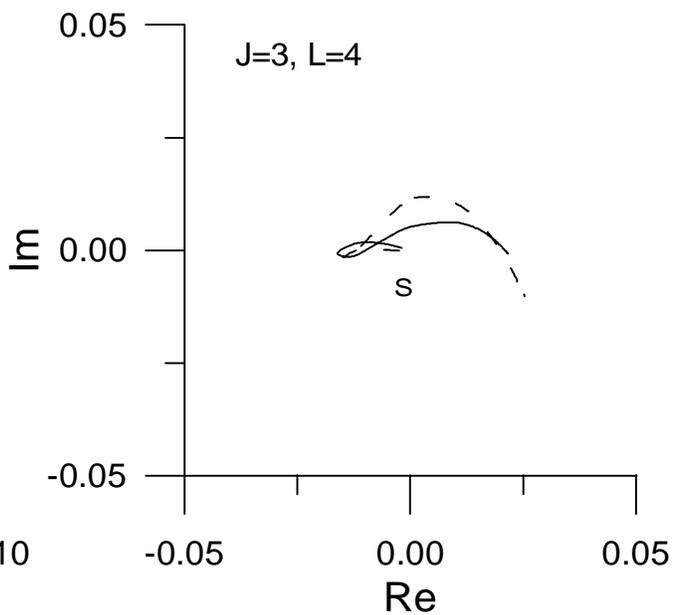
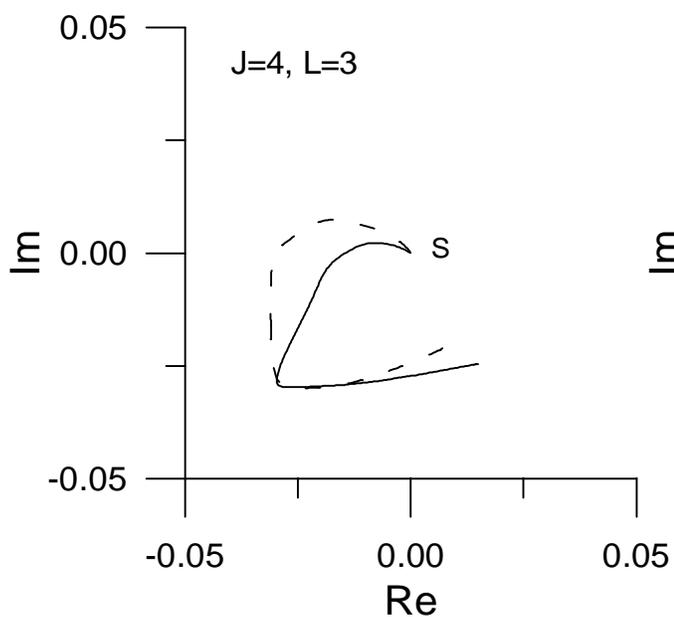
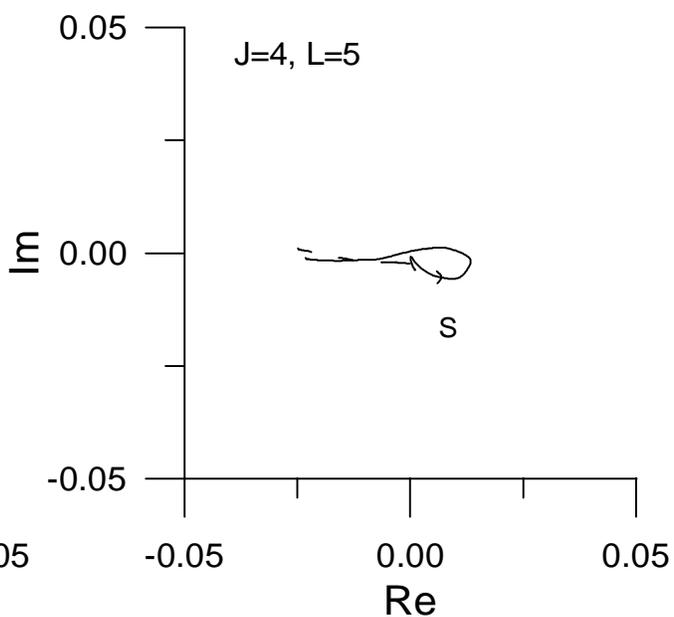
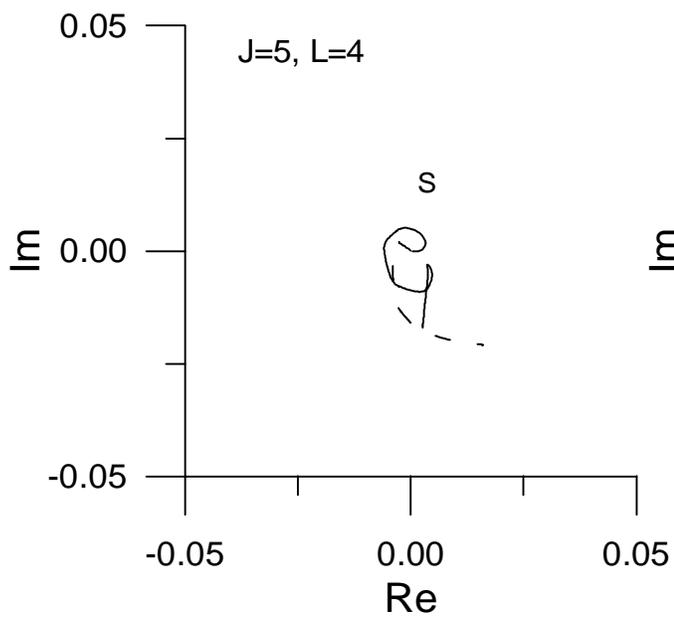
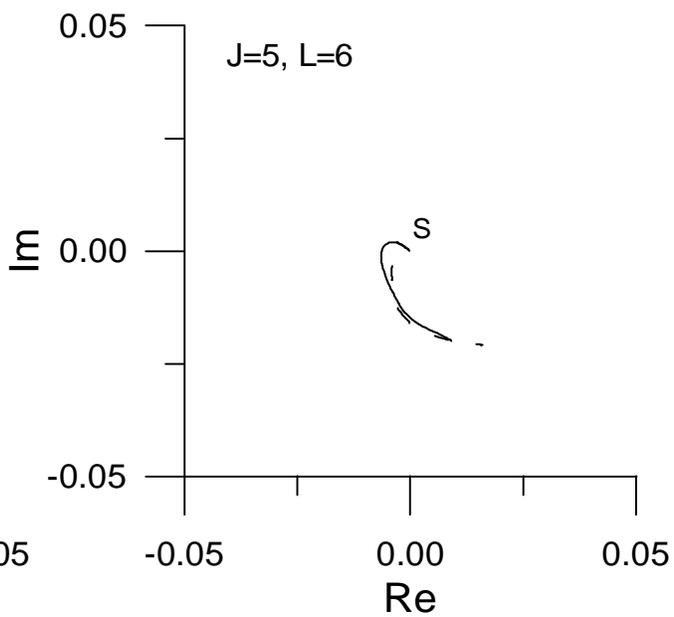
\end{document}